\documentstyle[pre,aps,multicol,epsfig,tabularx]{revtex}  
\begin{document}  
  
\draft  
  
\title{Size Segregation of Granular Matter in Silo Discharges}  
\author{Azadeh Samadani, A. Pradhan, and A. Kudrolli}  
\address{Department of Physics, Clark University, Worcester, MA  
01610, USA.}  
  
\date{\today} \maketitle  
  
\begin{abstract}  
  
We present an experimental study of segregation of granular matter in  
a quasi-two dimensional silo emptying out of an orifice. Size  
separation is observed when  
multi-sized particles are used with the larger particles found in the  
center of the silo in the region of fastest flow. We use imaging to  
study the flow inside the silo and quantitatively measure  
the concentration profiles of bi-disperse beads as a function of  
position and time. The angle of the surface is given by the angle of  
repose of the particles, and the flow occurs in a few layers  only  
near the top of this inclined surface. The flowing region becomes 
deeper  near the center of the silo and is confined to a  parabolic 
region centered at the orifice which is approximately  described by 
the kinematic model. The experimental evidence suggests that  the 
segregation occurs on the surface and not in the flow deep inside  
the silo where velocity gradients also are present. We report the  
time development of the concentrations of the bi-disperse particles  
as a function of size ratios, flow rate, and the ratio of initial  
mixture. The qualitative aspects of the observed phenomena may be  
explained by a void filling model of segregation. 

\end{abstract}  
  
  
\begin{multicols}{2}  
  
\narrowtext
 
\section{Introduction}  
  
Segregation is often observed in granular matter subject to shear or  
external excitation. However, there have been very few studies where  
quantitative information on the development of segregation is  
available. The nature of segregation depends on many factors such as  
the geometry and the surface properties of the particles, velocity  
gradients, and boundary  
conditions~\cite{stephens,bridgewater93,jaeger96}. For example in  
vibrated granular matter, segregation is observed, but the size  
concentration profiles depends on the shape of the container and the  
direction of the resulting  
convection~\cite{rosato87,knight96,knight97}. In partially filled  
rotating cylinders, axial segregation depends on factors such as the  
filling fraction and rotation rates~\cite{zik94,hill95,khakhar97}.  
The absence of a satisfactory continuum theory to describe the  
macroscopic properties of granular matter and the complicated nature  
of the complex convection patterns which result under the above  
conditions makes the analysis of the segregation difficult. 
 
In  comparison, gravity driven flows offer an  alternative for  
studying segregation where some progress has been made in 
understanding  the  flow \cite{savage88,nedderman79}. One of the 
simplest types of flow is the slow  flow of dense mixtures down an 
inclined surface. In this case free surface  segregation has been 
studied using bi-disperse  particles~\cite{savage88}. Preferential 
void filling of small  particles through the shear layer was 
identified as the main  mechanism of segregation. This flow and the 
resulting surface  segregation is similar to the situation where 
granular  matter is poured into a silo~\cite{cizeau99}. In this 
case, the angle  of repose quickly develops and the resulting flow 
is confined to a  few layers at the surface. For poly-disperse 
particles with  similar surface properties, the larger particles 
are found at the  bottom of the inclined surface and the smaller 
particles are found at  the top. This segregation can be understood 
in the same way as the free surface segregation and can also be 
understood in terms of a  capture 
model~\cite{boutreux96,boutreux98c} based on work by Bouchaud  et 
al.~\cite{bouchaud94,bouchaud95}. In the capture model the system 
is  divided into flowing and static regions and the smaller 
particles are assumed to  be more easily captured by the static 
layer than the larger ones. 
  
In this paper we consider flow in a silo which is filled uniformly 
with bi-disperse particles and then drained from  an orifice at the 
bottom. The nature of the flow is different because the flux of the 
particles is leaving the silo thus resulting in different boundary 
conditions. The resulting flow is more complex due to the 
development of a free surface and convergent flow near the  
orifice. Density waves are also known to occur~\cite{baxter89}. 
 
To  our knowledge, the only quantitative study of segregation in 
silo  discharges is by Arteaga and Tuzun~\cite{arteaga90} who 
measured the  volume ratio of a bi-disperse mixture as a function 
of time, but did  not visualize the internal flow. The ratio was 
found to be  independent of time except at the very end of the 
discharge. It was speculated that the development of velocity 
gradients in the  bulk due to the convergent flow is important for 
the observed  segregation. The questions we address here are (1) 
Where does the  segregation occur in the system? (2) What is the 
mechanism of  segregation, that is, does it occur due to 
preferential void filling  by small  particles at the surface or is 
it due to the development of velocity  gradients inside the silo? 
(3) What role does gravity play in  segregation? 
  
To address these questions, we used high resolution digital imaging  
to obtain detailed quantitative information on the evolution of  
particle  segregation inside a quasi-two-dimensional silo. We find  
that there are static regions where there is no flow  and mobile 
regions where the flow is very rapid. The latter region  is 
parabolic in shape about the orifice. We observe  that particles 
segregate so that the larger  particles are found in the  middle 
of the silo where the flow velocity is maximum. Although the  large 
particles are found in the region with the maximum velocity,  
segregation actually occurs at the surface. The extent of  
segregation depends on the size ratio and the relative number of  
large and small particles in the initial mixture. This observed  
phenomena is consistent with the void filling model of segregation  
developed in the context of flow down a rough inclined 
plane~\cite{savage88}. However, there are significant  differences 
in the extent of segregation because of the details of  the flow.  
  
\section{Experimental Apparatus}  
  
Figure~\ref{silo-fig} shows the schematic diagram of the experimental  
setup. A rectangular silo of dimensions $89\, {\rm cm} \times 45\,  
{\rm cm}$ and a width $w$ of 1.27 cm with an orifice at the bottom is  
used for the experiments. The surface and bulk flow inside the silo  
is visualized through the glass side-walls of the silo using a  
1000 $\times$ 1000 pixel non-interlaced Kodak ES 1.0 digital camera.  
A  
layer of $0.5\, {\rm mm}$ glass beads is glued to the bottom surface 
to  obtain non-slip boundary conditions. The shape of the orifice 
is  rectangular with dimensions $0.63\, {\rm cm}  
\times 1.27\, {\rm cm}$. A valve controls the flow rate through the  
orifice. The resulting flow is observed to be essentially  
two-dimensional. Limited experiments were also performed with a silo  
width of $2.54\, {\rm cm}$ to study the effect of the side walls on 
the flow.  
  
We use glass beads with various sizes and  
shapes as listed in Table I~\cite{jaygo}. The experiments are  
conducted in a controlled environment where the humidity is kept at  
15\% and the temperature is about~$38^\circ\, {\rm C}$. The  
data is insensitive to the temperature, but is sensitive to 
humidity  above 25\%.

\section{Observations}  
  
We first discuss the data corresponding to 0.5 mm mono-disperse glass  
beads (type M-1) to illustrate how the flow develops as a  
function of time. A sequence of images of the glass beads as they  
discharge from the silo is shown in Fig.~\ref{discharge}. The  
discharge rate  
$Q= 15\, {\rm g/s}$. The surface initially develops into an 
inverted  Gaussian shape as shown in Fig.~\ref{discharge}b, and no  
rolling of grains is observed at the surface. Soon after, rolling  
occurs after the local slope exceeds the angle of repose  
$\alpha$ for the grains. At the same time the surface develops  
a V-shape which is shown in Fig.~\ref{discharge}c. The angle the  
inclined surface makes with the horizontal equals the angle of  
repose and remains constant throughout the discharge, which takes  
approximately 500\,s for this case. The length of the inclined  
surface increases linearly until the top of the surface reaches 
the  side of the silo, and then remains constant.  
  
\subsection{Flow inside the Silo}  
  
To identify the regions in motion, we subtract two images separated  
by  
a short time interval. Figure~\ref{subtracted-img} shows the result  
of two images separated by 1.0~s. In the regions where the particles  
move, the intensities do not subtract to zero and give  
the speckled points in Fig.~\ref{subtracted-img}. Note that  
particles are completely stationary in the regions which are  
near the sides and a few layers from the surface. We divide the  
flowing phase into three regions as indicated in  
Fig.~\ref{subtracted-img}. The {\it surface flow} region, where the 
depth  of the mobile layers increases roughly linearly down the 
incline.  The {\it crossover} region, where the direction of the 
mean flow changes from along the surface to pointing toward the  
orifice. The  {\it internal} region, where the convergent flow  is 
very far from the surface.  The interface  between the moving and 
static regions~\cite{interface} is plotted in Fig.~\ref{compare} 
and is obtained  by averaging over two nearest neighbors to average 
out fluctuations.  
  
\medskip \noindent {\it Flow near the orifice}. The velocity  
distribution of the granular matter deep inside the silo has been  
described by a kinematic theory~\cite{nedderman79} based on a linear  
approximation for the relation between the horizontal component of  
the velocity  
$u$ and the gradient of the vertical velocity $dv/dx$, that is,  
\begin{equation} u = - B \times  
\frac{dv}{dx} \,. \label{kine-approx} \end{equation} where $B$ is a  
constant which has dimensions of length. Combining  
Eq.~(\ref{kine-approx}) with the continuity equation yields  
an equation for the vertical component of the velocity  
$v$~\cite{nedderman79}:  
\begin{equation}  
\frac{\partial v}{\partial y} = B \times \frac{\partial^2 v}{\partial  
x^2} \,. \label{grad} \end{equation}  
The solution to Eq.~(\ref{grad}) with appropriate boundary 
conditions  in the converging flow regime is  
\begin{equation} v(x,y) =  
\frac{Q}{\rho  
\sqrt{4 \pi B y}} e^{-x^2/4 B y} \,, \label{verticalv}  
\end{equation}  
where $x$ and $y$ measure the horizontal and  
vertical position from the orifice, and $\rho$ is the average mass  
density of particles relative to random close packing of the beads.  
Equation~(\ref{verticalv}) can be rewritten to give the  
equivelocity contours.  
\begin{equation} y = a(v) x^2 +2 B a(v)\,  y \ln y \, ,  
\label{equiv}  
\end{equation} where $a(v) =1/(4 B \ln (v\sqrt{4B\pi}\rho/Q))$. The 
stream lines are  also predicted by this theory and are given by 
\begin{equation} \psi  = \frac{Q}{2 \rho} {\rm erf}( 
\frac{x}{2\sqrt{4 B y}}) \,.  
\label{psi} \end{equation}  
Along the stream line $\psi$ is a 
constant, and Eq.~(\ref{psi})  gives \begin{equation} x = x_{i}  
\sqrt{B y} \, , \label{parabola} \end{equation}  
where $x_{i}$ is a constant that depends on the streamline.  
  
Predictions similar to Eqs.~(\ref{verticalv}) and (\ref{parabola}) 
have  been made using a diffusing void  
model~\cite{mullins74,caram91}. In this model voids are assumed to  
diffuse from the orifice to the surface. Using a biased random walk  
model for the motion of the voids, the mean flow velocity can be  
estimated by calculating the frequency of a walker visiting a  
lattice site inside the silo.

We compare our data to the prediction of kinematic theory with $B$  
as a fitting parameter in Fig.~\ref{compare}. The theory is  
in reasonable agreement with the data, especially considering the  
simplicity of the model. However, as can be seen in  
Fig.~\ref{compare},  if we consider only the first term in  
Eq.~\ref{equiv}, we obtain a better fit to the data. We  
summarize the results for the parameter, $a(v)$, and the  
kinematic model parameter, $B$, from fits to Eq.~\ref{equiv} in  
Table II. No significant dependence of the shape of the mobile 
region on the  flow rate and the width of system was  
found, but $B$ was found to depend on  
the size of the beads, which is consistent with the observations of  
Tuzun and Nedderman~\cite{tuzun79}.  
  
\medskip \noindent {\it Flow near the surface}. The  
description of flow using the kinematic model works only near the  
orifice. In the regions near the surface which are away from the  
sides, the flow is confined to a few layers. The number of layers  
increases linearly down the inclined surface to about 10 layers. As  
the silo empties, the surface moves down, and the static layers  
begin to flow. The depth dependence of the velocity at the point  
where 10 layers are moving is approximately linear, that is,  
\begin{equation} v_s = v_0 (1 - d /10)  
\quad \mbox{for}\quad d < 10 \, , \end{equation}  
where $v_0$ is the  
average velocity of the particles at the surface, and $d$ is the  
depth from the surface normalized by the mean particle size.  
  
If all particles which leave the surface flow region eventually  
leave the silo from the orifice, then the velocity can be  
approximately related to the flux of material leaving the silo.  
Therefore the vertical and horizontal component of the velocity  
at the surface can be written as:  
\begin{equation} v = \frac{Q}{10 d \rho} \sin(\alpha) \,\,\,\,  
\mbox{and} \, u = \frac{Q}{10 d \rho} \cos(\alpha) \,.  
\label{surface-v} \end{equation}  
Therefore the velocity of the particles in the crossover regime  
changes from that given by Eq.~(\ref{surface-v}) to that given by  
Eqs.~(\ref{kine-approx}) and~(\ref{verticalv}). The direction of the 
mean  flow changes from being parallel to the surface to pointing 
down  towards the orifice. This crossover region is approximately 
$15\, {\rm cm}  
\times 15\, {\rm cm}$ corresponding to hundreds of grains. We 
discuss in  Sec.~IV the details of the flow in the crossover regime 
and its effect on the observed segregation.  
  
We note that the flow near the  
surface and around the obstacles has been also modeled by the 
diffusing   void model~\cite{caram91}. This  
model gives a crossover which is very sharp and is of the order of  
one grain diameter. The scale over which the crossover occurs in   
our experiments is much broader as can be seen in  
Fig.~\ref{compare}.  
  
\subsection{Segregation}  
  
With this description of flow inside a discharging silo, we now  
report our experiments on bi-disperse glass particles to study  
segregation. To visualize the segregation, two sizes of beads with  
different colors but identical surface properties are used. The  
symbol $P_W$ is used to specify the percentage of larger particles  
by weight in the initial mixture in the silo. We first  discuss the 
development of segregation when 1.2 mm yellow glass 
particles (M-4) are mixed with 0.6 mm blue glass  
particles (M-2) and therefore the diameter ratio is $r = 2$.  
  
We pour the mixture ($P_W = 10\%$) into the silo as  
uniformly as possible. The development of the flow is similar to the  
mono-disperse case described earlier, but we also observe that the  
density of the larger grains increases near the surface and the  
density of smaller grains increases in the moving layers below the  
surface. The evolution of segregation is  
shown in Fig.~\ref{segregation}.  
  
To parameterize the segregation we measure the ratio of the two  
types of beads in a horizontal narrow rectangular region at a  
height of 5\,cm above the orifice by  measuring the light  
intensity. The light intensity is a monotonic  function of the  
density ratio of the two kind of particles. This  function is  
determined by using known weight ratios of particles in a  separate  
series of calibration experiments. The density ratio of the   
particles is plotted as a function of horizontal  position and time  
in Fig.~\ref{seg-den}.  We observe from Fig.~\ref{seg-den} that the  
density of larger particles  increases in the midpoint (directly  
above the orifice) as a function of  time. The percentage of larger  
particles at the midpoint increases from 10\%, corresponding to the  
initial mixture, to about 100\%.  
  
To further characterize the evolution of the segregation, we have  
plotted  the density fraction of the large beads at the center point  
as a function of time (see Fig.~\ref{peak-seg}(a)). This density is  
called the ``segregation parameter'' $s(t)$. From  
Fig.~\ref{peak-seg}(a) it can be seen that there is  no segregation  
for the first 50\,s after the start of the flow as can be also seen  
from Fig.~\ref{seg-den}. During  this time we note that the  
velocity gradients of the grains in the silo are fully developed.  
The fact that $s(t)$ increases only after $t=50$\, s indicates 
that  the velocity gradients deep inside the silo are not 
responsible for  segregation.  We observe that a thin band of 
larger  particles initially appears at the surface and grows  
down towards the orifice (see  Fig.~\ref{segregation}.) Because 
the observed area is near  the bottom of the silo (5\,cm  
above the orifice), it takes about 50\,s for the larger particles 
to  travel to the measured region for a  flow rate of 15.0\,g/s. 
After 50\,s,  
$s(t)$ increases to about 1 and remains constant.  
  
\medskip \noindent{\it Effect of size ratio}. We repeated the  
experiment with different size ratios to investigate the effect 
of different sizes on  the segregation rates. For a mixture of 
0.6\,mm  black glass beads (M-2) and 0.7\,mm red glass beads (M-3) 
with  
$P_W=15\%$, the  size ratio $r \approx 1.2$. Segregation is observed 
even for such a small size difference. A separate set of  
calibration curves for the density  ratio of the particles was  
obtained in this case. The  density distribution as a  
function of position and time with a flow  rate of 15\,g/s is  
similar to the larger size ratio, but the segregation does not  
occurs as quickly and is less in comparison to the larger size  
ratio. As seen in Fig.~\ref{peak-seg}(b), $s(t) \approx 0.15$  for 
$t \leq 50\, {\rm s}$, and then grows to about 0.3 -- 0.35.  
Therefore  
$s(t)$ for $r=1.2$ is substantially smaller than for $r=2.0$.  
  
For $r=1.2$, not only is the extent of segregation  
lower, but the interface between the region containing the large  
and small particles is more diffuse. The mass density ratio of the  
larger beads in the narrow rectangular region under observation  
is shown in Fig.~\ref{ratio} for $r = 2.0$ and  
$r = 1.2$. Both mass density ratios may be fitted by a Gaussian,  
with the Gaussian for higher $r$ substantially narrower. In  
Section~IV, we argue that the diffused nature of the interface  
between large and small particles for $r=1.2$ is a result of the  
nature of the flow in the crossover region.  
  
It can be also seen from Fig.~\ref{peak-seg} that the saturation  
point of the segregation parameter $s(t)$ depends on the diameter  
ratio  
$r$ and the density ratio $P_W$. If the segregation is not  
complete, there are fluctuations in the number ratio of the  
particles about the saturation value. The fluctuations are stronger  
when the saturation is lower.  
  
Experiments were also performed with polydisperse particles (P-1 and  
P-2 in Table~I), and segregation was also observed with larger  
particles found at the center as in the bi-disperse cases already  
discussed. No quantitative data was obtained for polydisperse  
particles.  
  
\medskip \noindent{\it Effect of number ratio of large and small  
particles.}  
We considered the effect of smaller values of the density 
ratio $P_W$  on 
the extent of segregation by  doing experiments with $P_W=30\%$ and 
50\%. The results of  these experiments also are  plotted in 
Fig.~\ref{peak-seg}(b). The  overall development of segregation is 
similar, but the extent  depends on $P_W$. From  
Fig.~\ref{peak-seg}(b), we observe that the saturation level of  
segregation  increases for higher $P_W$. The saturation value is  
found to fluctuate on the order of 5\%.  
  
\medskip \noindent{\it Effect of flow rate.}  
We repeated all the experiments with a flow rate of $3.0 \pm 0.5\, 
{\rm g/s}$, which corresponds to the slowest rate for which  
continuous  flow is possible in our system. The behavior of the 
segregation  parameter $s(t)$ is similar to the faster flow rate of 
15\, g/s and  is shown in Fig.~\ref{peak-seg}(c).  For the slower 
flow rate, no  segregation occurs for $t < 200$\,s. This time is 
longer because  the segregated particles at the surface take a 
longer time to arrive  in the region where we monitor $s(t)$. The 
ratio of times for the  development of the segregation is 
approximately the same as the  ratio of the flow rates. There are 
small changes in the value of  
$s(t)$, but these changes are the same order as the errors in the  
calibration of the density ratio. 
  
\section{Discussion}  
  
Because the data clearly suggests that segregation occurs at  the 
surface (see Fig.~\ref{seg-pic}),  we first explore the relevance of a  model proposed by 
Savage and Lun~\cite{savage88} based on their  experiments on 
simple flows on an inclined plane. They considered a  shear flow of 
a thin layer of bi-disperse glass beads down a rough  inclined 
plane and obtained quantitative data for  the development of 
segregation. The beads  were collected from different  heights 
in the layer using an arrangement of baffles which directed  the 
beads into different bins~\cite{savage88}. They  considered two 
mechanisms to explain the development of  segregation: (i) the 
preferential filling of voids in a lower layer  by smaller 
particles, and (ii) the expulsion of particles to the top  layer 
(which is not size dependent) to make the net flux through a  layer 
zero.  
  
The probability of inter-layer percolation of particles is  
calculated by considering the relative probabilities of small and 
large particles  falling into possible voids as  a function of the 
size and the number ratio of  the two types of particles 
(see Fig.~\ref{model}). The resulting  probability is an exponential 
function of particle sizes and  average void size~\cite{savage88}. 
Therefore, the probability for  small particles to fall into voids 
is significantly higher than  that for larger particles if the 
voids created are of a similar  size. By then assuming a linear 
profile for the velocity as a  function of depth in the layer and a 
constant velocity along the  inclined plane, they were able to 
calculate relations for the  concentration profiles of the 
particles as a function of depth in  the layer and position down 
the inclined plane. Their model  predicts that the particles 
segregate completely after traveling a  certain distance down the 
incline which depends on the size ratio and volume density ratio of 
the initial mixture,  and the angle of inclination of the plane. 
  
In our experiments in silo discharges, the flow in the 
surface flow region (see Fig.~\ref{subtracted-img}) is similar to 
that considered by  
Savage and Lun~\cite{savage88}. Therefore, we can qualitatively 
explain the  observed segregation of particles inside the silo by 
the  mechanism of preferential filling of voids by smaller 
particles  near the surface.  Particles are then carried along the  
streamlines which results in the larger particles being in the  
central region of the silo where the flow velocity is highest (also 
see Fig.~\ref{seg-pic}).  
However, a quantitative comparison with the predictions of  
Ref.~\cite{savage88} cannot be done because of significant  
differences in  
the underlying flow. In our experiments, the flux of  
particles comes from the static layers being converted to mobile  
layers, whereas in the simple inclined flow, the particles enter at  
one point at the top of the inclined plane. In addition, the flow  
near the bottom of the surface acquires a significant vertical  
component.  
  
We also find that the inclined surface grows linearly as the silo 
discharges  until  the surface reaches the edge of the silo and 
then remains  constant. If the segregation depends on the length of 
the surface  as calculated in Ref.~\cite{savage88}, then we would 
expect $s(t)$ to  increase during the time the surface length 
increases. In our  experiments, $s(t)$ saturates to values less 
than 100\% for small  
$r$ over a range of values of  
$P_W$ during a time which is less than the time for which the  
surface length increases. The complications introduced by  the 
additional features in the flow have to be taken into account to  
explain saturation of less than 100\% in $s(t)$ for small $r$. 
  
One possible explanation for the saturation is as follows.  In 
addition to voids being created due to fluctuations in the  
inter-layer velocity, the vertical component of the velocity 
becomes  significant in the crossover regime. Thus particles create 
larger  voids behind them in the crossover regime (see 
Fig.~\ref{subtracted-img}) in comparison  to the surface flow regime. 
Therefore, the probability of finding  larger voids where large 
particles can fall into increases in the  crossover regime. At some 
point on the inclined surface, the  difference between the 
probability of a void being filled by a small  or large particle 
becomes negligible. This point might be expected to occur  higher on 
the inclined plane for smaller size differences resulting in  lower 
saturation values for $s(t)$.  
 
This saturation appears to occur for $r = 1.2$ as seen in  
Fig.~\ref{ratio}. If the size ratio $r$ is large enough, the  
probability for smaller  
particles to fill a void remains much higher and  
complete segregation is seen for $r=2$ in Fig.~\ref{ratio}.  
  
It might be expected that the change in flow rate would affect  
segregation because the creation of voids depends on the  
fluctuations and value of the mean velocity. Because the void  
filling mechanism is important for  
segregation, the quantitative progress of segregation may   
depend on the size ratio and flow rate. However, these changes  
appear to be not important in the range of flow rate available in  
our experiments.  
 
The data for $t < 50$\,s and $Q = 15$\,g/s in Fig.~\ref{peak-seg} 
also  shows that the void filling mechanism is unimportant if the  
direction of flow is in the same direction as the gravitational  
field as is the case in the deep flow regime of the silo.  
Segregation of bi-disperse particles in the absence of a  
gravitational field has been considered in  a ``collisional" flow by  
Jenkins~\cite{jenkins98}. The difference in the scattering rates  
for the two different size particles is anticipated to give rise  
to segregation for rapid flow. However, the velocities  
of the particles in the silo may be too small for such effects to be  
important.  
  
\section{Summary}  
  
In summary, we have reported experiments on segregation of  
bi-mixtures of glass beads in a discharging silo. Using digital  
imaging we characterized the flow and measured the development of  
segregation as a function of position and time. The flow deep  
inside the silo is approximately characterized by the kinematic  
model~\cite{nedderman79}, but more theoretical developments are  
required to  
model the flow near the surface. Segregation is observed even for  
very small size ratios. We observe that segregation occurs at the  
surface and not in the bulk where velocity gradients are also  
present. We quantitatively characterize the development and  
progress of the segregation using the mass ratio $s(t)$. We also  
obtained quantitative information about the size distribution of the  
particles inside the silo (see Fig.~\ref{ratio} for example). Our  
experiments also indicate that the segregation progresses very  
quickly if the surface flow is not along the direction of the  
gravitational field.  
  
A qualitative explanation of the distribution of concentration of  
large and small particles can be given by using the void filling  
mechanism. However, a quantitative explanation requires a better  
understanding of the velocity profiles near the surface and its role  
on creating voids which drive flow and segregation. Quantitative  
experimental data for the spatial and time development of segregation  
is scarce, and the data presented in this paper provides a  
guide to the development of models of segregation.  
  
We thank D. Hong, L. Mahadevan and H. Gould for useful discussions,  
and J. Norton for technical assistance. This work was partially supported 
by the the Donors of the Petroleum Research Fund administered  
by the American Chemical Society and one of us (A. K.) was also  
funded by the Alfred P.\ Sloan Foundation.

\begin{figure}  
\caption{Schematic diagram of the experimental apparatus. Grains pour  
out of an orifice at the bottom of a quasi-two-dimensional silo with  
a flow rate $Q$. The dashed box ($30\,{\rm cm} \times 30\,{\rm cm}$) 
indicates  the area corresponding to images in 
Figs.~\ref{discharge}  and~\ref{segregation}.}  
\label{silo-fig}  
\end{figure}  
  
  
\begin{figure}   
\caption{Sequence of images of 0.5 mm monodisperse glass beads  
(type M-1) discharging ($Q = 15$\,g/s) as a function of time (a): 
$t =  10$\,s, (b): $t = 20$\,s, (c): $t = 100$\,s, (d): $t = 
200$\,s.}  
\label{discharge} 
\end{figure}  
  
\begin{figure}
\caption{Difference of images separated by time difference of  
$1\, {\rm s}$. The regions which appear as gray are in motion ($Q =  
15$\, g/s.) The regions in motion can be divided into the {\em surface 
flow} 
region, the {\em crossover} region and the {\em internal} region.} 
\label{subtracted-img}  
\end{figure}  
  
  
\begin{figure}   
\caption{The interface between the static and mobile regions at  
time $t=100\,s$ ($\Box$) and $t = 200$\,s ($\bullet$). The  
interface actually corresponds to an equivelocity curve for  
which $v = 0.03\,{\rm cm/s}$. Also shown is the comparison of the  
interface to a fit    
of the kinematic model (Eq.~(\protect{\ref{equiv}})) (dashed line) 
and a fit to a  parabola (solid line). Note that far from the 
orifice the  observed interface  crosses over to a line which is 
parallel to the  surface.}  
\label{compare}  
\end{figure}  
  
\begin{figure}  
\caption{Images  of bi-disperse particles inside the silo as a 
function of time ($r =  2.0$; $Q = 15$\, g/s.) The larger particles 
are found in the center 
of  the flow. } 
\label{segregation} 
\end{figure}  
  
\begin{figure}  
\caption{The  
fraction of large particles as a function of time in a narrow 
region  located at a height of 5\,cm above the orifice. A peak 
develops at the  center which grows higher and wider corresponding 
to the segregation  of particles ($r = 2.0$, $P_W = 10\%$, $Q = 
15$\,g/s.) The dashed  vertical lines indicate the mobile region.} 
\label{seg-den}  
\end{figure}  
  
\begin{figure}  
\caption{(a) The segregation parameter $s(t)$ for $r = 2$, and $P_W 
= 10\%$, and $Q =  15$\, g/s. (b) The segregation for different 
initial values of $P_W$ for $r=1.2$. (c) Same as the data in (b) but 
at a lower flow rate ($Q =  3.0$\,g/s.)} 
\label{peak-seg} 
\end{figure}  
  
\begin{figure}  
\caption{The fraction of larger particles for $ r = 1.2$ 
($\bullet$) and $r =  2.0$ ($\circ$) as function of position at a 
height of 5\,cm above the  orifice at  
($t = 250$\,s). The data is also fitted with a Gaussian to guide 
the  eye. The region  
between large and small particles is more diffused for lower $r$.}  
\label{ratio} 
\end{figure}  
  
\begin{figure}  
\caption{An image of segregated particles. The larger particles  
(black) are rolling on  
top of layers of small particles (white) ($r = 2$.)} 
\label{seg-pic}  
\end{figure}  
  
\begin{figure}  
\caption{The void filling mechanism at the surface appears to be  
responsible for the observed segregation. Smaller particles have a 
higher  probability to fill a void compared to larger particles 
for a  comparable sized void.}   
\label{model}  
\end{figure}  
  
\begin{table}
\begin{center} 
\begin{tabular}{l c c c} \hline  Type \, &  
\, Distribution \, & \, Size (mm) \, & \, Angle of Repose, $\alpha$ 
\, \\ \hline  
\hline M-1 & Mono-disperse &0.5 $\pm$ 0.05  & $24.5^\circ \pm  
0.2^\circ$  
\\ M-2 & Mono-disperse  &0.6 $\pm$ 0.05  & $24.7^\circ\pm  
0.2^\circ$ \\ M-3  & Mono-disperse &0.7 $\pm$ 0.05 &  
$24.5^\circ\pm0.2^\circ$ \\ M-4 & Mono-disperse &1.1 $\pm$ 0.10  &  
$24.2^\circ\pm0.2^\circ$ \\ P-1 & Poly-disperse  &0.5 $ - $ 0.05  &  
$25.6^\circ \pm 0.2^\circ$  
\\ P-2  & Poly-disperse  &0.2 $ - $ 0.05 & $25.6^\circ \pm  
0.2^\circ$\\  
\hline 
\end{tabular} 
\end{center} 
\caption{Types of glass particles used in the experiments. The particles are approximately 
spherical.}  
\end{table}  

\begin{table}  
\begin{center}  
\begin{tabular}{l c c c}  
\hline  
  
& $w = 1.27$\,cm & $w = 2.54$\,cm & $w = 1.27$\, cm \\   
Type & $a(v)$ (cm) & $a(v)$ (cm)   & $B$ (cm) \\ \hline \hline   
M-4 &$0.20 \pm 0.00$ &$0.20  \pm 0.00$ & $0.38 \pm 0.04$ \\   
M-2 &$0.32 \pm 0.01$ &$0.36 \pm  0.01$ & $0.21 \pm 0.02$ \\   
P-1 &$0.40 \pm 0.03$ &$0.40  \pm 0.03$ & $0.18 \pm 0.02$ \\   
P-2 &$0.40 \pm 0.03$ & - & - \\ \hline  
  
\end{tabular} 
\end{center} 
\caption{The width $a(v)$ (see  
Eq.~\ref{equiv}) of the parabolic fit that describes the interface of  
the deep flow near the orifice for different size beads and width  
$w$ of the silo ($v = 0.03$\, cm/s.) The parameter $B$ which relates the horizontal  
component of the velocity to the gradient of the vertical  
component in the kinematic model (see Eq.~\ref{equiv}) also  
describes the interface. The flow rate  
$Q= 4.5\,{\rm g/s}$.}  
\end{table}

\end{multicols}


\begin{thebibliography}{}
\bibitem{interface} This interface corresponds to an equivelocity 
contour  for a very small magnitude of velocity which is of the 
order of the size of a  pixel divided by the time difference 
between the two images used in  identifying the moving regions.
\end{thebibliography}
\end{document}